\documentclass[12pt,english,aps, final]{revtex4}
\usepackage[T1]{fontenc}
\usepackage[latin9]{inputenc}
\usepackage{array}
\usepackage{multirow}
\usepackage{amsmath}
\usepackage{amssymb}
\usepackage{esint}

\makeatletter

\providecommand{\tabularnewline}{\\}

\@ifundefined{textcolor}{}
{%
 \definecolor{BLACK}{gray}{0}
 \definecolor{WHITE}{gray}{1}
 \definecolor{RED}{rgb}{1,0,0}
 \definecolor{GREEN}{rgb}{0,1,0}
 \definecolor{BLUE}{rgb}{0,0,1}
 \definecolor{CYAN}{cmyk}{1,0,0,0}
 \definecolor{MAGENTA}{cmyk}{0,1,0,0}
 \definecolor{YELLOW}{cmyk}{0,0,1,0}
}

\makeatother

\usepackage{babel}
\begin{document}
\title{Dimensional Regularization and Dispersive Two-Loop Calculations}
\author{A. Aleksejevs and S. Barkanova}
\affiliation{Grenfell Campus of Memorial University, Corner Brook, NL, Canada}
\begin{abstract}
The two-loop contributions are now often required by the precision
experiments, yet are hard to express analytically while keeping precision.
One way to approach this challenging task is via the dispersive approach,
allowing to replace sub-loop diagram by effective propagator. This
paper builds on our previous work, where we developed a general approach
based on representation of many-point Passarino-Veltman functions
in two-point function basis. In this work, we have extracted the UV-divergent
poles of the Passarino-Veltman functions analytically and presented
them as the dimensionally-regularized and multiply-subtracted dispersive
sub-loop insertions, including self-energy, triangle, box and pentagon
type.
\end{abstract}
\maketitle

\section{Introduction}

The electroweak precision searches for the physics beyond the Standard
Model (BSM) frequently demand a sub-percent level of accuracy from
both experiment and theory. For the new-generation precision experiments
such as MOLLER \citep{MOLLER} and P2 \citep{P2} , for example, that
means evaluating electroweak radiative corrections up to two-loop
level with massive propagators and control of kinematics, which is
a highly challenging task. In some cases, it may not possible to express
the final results analytically, so one would have to use approximations
and/or numerical methods. See, for example, an overview of numerical
loop integration techniques in \citep{Freitas1}, a general case of
the two-loop two-point function for arbitrary masses in \citep{Kreimer},
and a method of calculating scalar propagator and vertex functions
based on a double integral representation in \citep{Czarnecki} and
\citep{Frink}. The more recent developments on analytical evaluation
of two-loop self-energies can be found in \citep{Adams1,Adams2,Adams3,Remiddi1,Bloch1,Bloch2},
and on numerical evaluation of general n-point two-loop integrals
using sector decomposition in \citep{Borowka1,Borowka2}. The idea
of the sub-loop insertions with the help of the dispersive approach
was implemented for the self-energies \citep{Bohm}, \citep{Hollik-1}
and partially for the vertex graphs with the help of Feynman parametrization
\citep{Hollik-2}. A somewhat relevant case of the self-energy dispersive
insertions for Bhabha scattering in QED was considered in \citep{Gluza2005}
and \citep{Gluza2008}.

In \citep{AA1,AA2}, we have developed a general approach in calculations
of the two-loops diagrams, which is based on the representation of
many-point Passarino-Veltman (PV) functions in two-point function
basis. As a result, we where able to replace a sub-loop integral by
the dispersive representation of the two-point function. In that case,
the second loop received an additional propagator and we where able
to use the PV basis for the second loop integration in the final stage
of the calculations. The final results where presented in a compact
analytic form suitable for numerical evaluation. Since in the majority
of applications such two-loops integrals are either ultraviolet or
infrared (IR) divergent, a regularization scheme is required. In case
of the IR-divergence, the regularization can be done by introducing
a small mass of the photon which is later removed by a contribution
of a combination of one-photon bremsstrahlung from one-loop and two-photon
bremsstrahlung from tree level diagrams. Since the IR-divergence does
not impact convergence of the dispersion sub-loop integral, the mass
of the photon in the insertion could be carried into second loop without
an additional complications. If necessary, the dependence on the photon
mass can be extracted analytically. For the UV-divergent two-loops
diagrams, the regularization of the sub-loop insertion is done by
an introduction of a cut-off parameter for the divergent dispersive
integral. The second-loop regularization is done by dimensional regularization,
but in this case, when counter terms are added, one set of renormalization
constants is evaluated in dispersive approach with a cut-off parameter,
and another set of the constants is calculated using dimensional regularization.
In this case, the independence of the final results from the regularization
parameters could be confirmed numerically only. That can result in
additional complications, since the two-loops integrals could suffer
from a number of the numerical instabilities. In some simple cases,
when sub-loop renormalization is possible (for ex. box diagram with
self-energy insertion), one can represent the sub-loop by doubly-subtracted
dispersive integrals and carry on the second-loop integration using
the PV-function basis without dealing with additional UV divergences.
In this paper, we follow a general approach developed in \citep{AA1}
and extract the UV-divergent parts of the two-loop integrals analytically.
For that, we need to represent the UV-divergent dispersive sub-loop
insertion using dimensional regularization and extract UV poles analytically.
Since in \citep{AA1,AA2} the two-loop integrals where all reduced
to the two-point PV-function basis, we start with the outline of the
ideas on how to express the two-point sub-loop insertion with UV divergent
part written out in the dimensional regularization and the UV-finite
part represented by a multiply-subtracted dispersive integral. Later,
we extend this approach to triangle-, box- and pentagon-type of insertions. 

\section{Methodology}

Generally, a two-point function of an arbitrary rank could be written
in the dimensional regularization as: 
\begin{align}
{\displaystyle B_{\underset{2l}{\underbrace{0...0}}\underset{n}{\underbrace{1...1}}}\left(p^{2},m_{1}^{2},m_{2}^{2}\right)} & \equiv B_{\{2l,n\}}\left(p^{2},m_{1}^{2},m_{2}^{2}\right)=\mu^{2\epsilon}e^{\gamma_{E}\epsilon}\frac{\left(-1\right)^{2+n+l}}{2^{l}}\Gamma\left(\epsilon-l\right)\nonumber \\
\label{eq:1}\\
 & \times\lim_{\varepsilon\rightarrow0^{+}}\intop_{0}^{1}dx\,x^{n}\left(p^{2}x^{2}+m_{1}^{2}+x\left(m_{2}^{2}-m_{1}^{2}-p^{2}\right)-i\varepsilon\right)^{-\epsilon+l}\nonumber 
\end{align}
Here, $\epsilon=\frac{4-D}{2}$ is the dimensional regularization
and $\mu$ is the mass-scale parameter. The UV-divergent part Eq.\ref{eq:1}
can be expressed as a polynomial in $p^{2}$ multiplied by $\left(\frac{1}{\epsilon}+\ln\frac{\mu^{2}}{m_{2}^{2}}\right)$
term. A linear term in $\epsilon$ will give rise to the local terms
after taking the second-loop integration, and can be considered as
a finite part of the two-point functions which has dependence on $\ln\frac{\mu^{2}}{m_{2}^{2}}$.
Hence, the regularized one-loop UV-divergent part has the following
form:
\begin{align}
B_{\{2l,n\}}^{UV}\left(p^{2},m_{1}^{2},m_{2}^{2}\right) & =\left(\frac{1}{\epsilon}+\ln\frac{\mu^{2}}{m_{2}^{2}}\right)\sum_{i=0}^{l}a_{i}^{\{2l,n\}}p^{2i}.\label{eq:2}
\end{align}
Here, coefficients $a_{i}^{\{2l,n\}}$ are the functions of masses
$m_{\{1,2\}}^{2}$ with structure provided in Tbl.\ref{tbl:1}.
\begin{table}
\begin{centering}
\begin{tabular}{|c|c|c|c|}
\hline 
$a_{i}^{\{2l,n\}}$ & $l=0$ & $l=1$ & $l=2$\tabularnewline
\hline 
\hline 
$n=0$ & $a_{0}^{\{0,0\}}=1$ & \multirow{1}{*}{$\begin{array}{cc}
a_{0}^{\{2,0\}}= & \frac{1}{4}\left(m_{1}^{2}+m_{2}^{2}\right)\\
a_{1}^{\{2,0\}}= & -\frac{1}{12}
\end{array}$} & $\begin{array}{cc}
a_{0}^{\{4,0\}}= & \frac{1}{24}\left(m_{1}^{4}+m_{2}^{4}+m_{1}^{2}m_{2}^{2}\right)\\
a_{1}^{\{4,0\}}= & -\frac{1}{48}\left(m_{1}^{2}+m_{2}^{2}\right)\\
a_{2}^{\{4,0\}}= & \frac{1}{240}
\end{array}$\tabularnewline
\hline 
$n=1$ & $a_{0}^{\{0,1\}}=-\frac{1}{2}$ & $\begin{array}{cc}
a_{0}^{\{2,1\}}= & -\frac{1}{12}\left(m_{1}^{2}+2m_{2}^{2}\right)\\
a_{1}^{\{2,1\}}= & \frac{1}{24}
\end{array}$ & $\begin{array}{cc}
a_{0}^{\{4,1\}}=- & \frac{1}{96}\left(m_{1}^{4}+3m_{2}^{4}+2m_{1}^{2}m_{2}^{2}\right)\\
a_{1}^{\{4,1\}}= & \frac{1}{240}\left(2m_{1}^{2}+3m_{2}^{2}\right)\\
a_{2}^{\{4,1\}}= & -\frac{1}{480}
\end{array}$\tabularnewline
\hline 
$n=2$ & $a_{0}^{\{0,2\}}=\frac{1}{3}$ & $\begin{array}{cc}
a_{0}^{\{2,2\}}= & \frac{1}{24}\left(m_{1}^{2}+3m_{2}^{2}\right)\\
a_{1}^{\{2,2\}}= & -\frac{1}{40}
\end{array}$ & $\begin{array}{cc}
a_{0}^{\{4,2\}}= & \frac{1}{240}\left(m_{1}^{4}+6m_{2}^{4}+3m_{1}^{2}m_{2}^{2}\right)\\
a_{1}^{\{4,2\}}= & -\frac{1}{240}\left(m_{1}^{2}+2m_{2}^{2}\right)\\
a_{2}^{\{4,2\}}= & \frac{1}{840}
\end{array}$\tabularnewline
\hline 
$n=3$ & $a_{0}^{\{0,3\}}=-\frac{1}{4}$ & $\begin{array}{cc}
a_{0}^{\{2,3\}}= & -\frac{1}{40}\left(m_{1}^{2}+4m_{2}^{2}\right)\\
a_{1}^{\{2,3\}}= & \frac{1}{60}
\end{array}$ & $\begin{array}{cc}
a_{0}^{\{4,3\}}= & -\frac{1}{480}\left(m_{1}^{4}+10m_{2}^{4}+4m_{1}^{2}m_{2}^{2}\right)\\
a_{1}^{\{4,3\}}= & \frac{1}{840}\left(2m_{1}^{2}+5m_{2}^{2}\right)\\
a_{2}^{\{4,3\}}= & -\frac{1}{1344}
\end{array}$\tabularnewline
\hline 
$n=4$ & $a_{0}^{\{0,4\}}=\frac{1}{5}$ & $\begin{array}{cc}
a_{0}^{\{2,4\}}= & \frac{1}{60}\left(m_{1}^{2}+5m_{2}^{2}\right)\\
a_{1}^{\{2,4\}}= & -\frac{1}{84}
\end{array}$ & $\begin{array}{cc}
a_{0}^{\{4,4\}}= & \frac{1}{840}\left(m_{1}^{4}+15m_{2}^{4}+5m_{1}^{2}m_{2}^{2}\right)\\
a_{1}^{\{4,4\}}= & -\frac{1}{672}\left(m_{1}^{2}+3m_{2}^{2}\right)\\
a_{2}^{\{4,4\}}= & \frac{1}{2016}
\end{array}$\tabularnewline
\hline 
\end{tabular}
\par\end{centering}
\caption{Coefficients $a_{i}^{\{2l,n\}}$ for $B_{\{2l,n\}}^{UV}$. }

\label{tbl:1}
\end{table}
In order to satisfy the definition given in Eq.\ref{eq:1}, the UV-divergent
pole $1/\epsilon$ in Eq.\ref{eq:2} should be treated as $\frac{1}{\epsilon}\rightarrow\frac{1}{\epsilon}-\gamma_{E}+\ln\left(4\pi\right)$.
In the case of sub-loop insertion, the UV part represented by Eq.\ref{eq:2}
can be easily carried into the second-loop integral. Here, the momentum
$p^{2}$ could depend on the momentum of the second loop and Feynman
parameters used in \citep{AA1}. In order to keep the UV-divergent
term presented in Eq.\ref{eq:2} as simple as possible, we will treat
masses as constants. In the case where masses depend on the Feynman
and mass shift parameters (see \citep{AA1}), a simple transformation
$\ln\frac{\mu^{2}}{m_{2}^{2}}\rightarrow\ln\frac{\mu^{2}}{m_{0}^{2}}+\ln\frac{m_{0}^{2}}{m_{2}^{2}}$
can be used, where $m_{0}$ is the arbitrary constant mass. A term
proportional to $\ln\frac{m_{0}^{2}}{m_{2}^{2}}$ is UV-finite and
scale-parameter independent, and hence can be moved to the UV-finite
part of Eq.\ref{eq:1} for which we will construct a dispersive representation.
The UV-finite part could be presented through the dispersive integral:
\begin{align}
B_{\{2l,n\}}^{fin}\left(p^{2},m_{1}^{2},m_{2}^{2}\right) & =\frac{1}{\pi}\intop_{\left(m_{1}+m_{2}\right)^{2}}^{\infty}ds\frac{\Im B_{\{2l,n\}}^{fin}\left(s,m_{1}^{2},m_{2}^{2}\right)}{s-p^{2}-i\varepsilon}.\label{eq:3}
\end{align}
Here, $B_{\{2l,n\}}^{fin}\left(p^{2},m_{1}^{2},m_{2}^{2}\right)$
is the UV-finite part of Eq.\ref{eq:1}: $B_{\{2l,n\}}\left(p^{2},m_{1}^{2},m_{2}^{2}\right)=B_{\{2l,n\}}^{UV}\left(p^{2},m_{1}^{2},m_{2}^{2}\right)+B_{\{2l,n\}}^{fin}\left(p^{2},m_{1}^{2},m_{2}^{2}\right)$.
The function $B_{\{2l,n\}}^{fin}\left(p^{2},m_{1}^{2},m_{2}^{2}\right)$
consists of the finite part of the two-point function, $b_{\{2l,n\}}^{fin}\left(p^{2},m_{1}^{2},m_{2}^{2}\right)$,
which is free from any of the regularization parameters plus an additional
terms linear in $\epsilon$, which are also finite. More specifically,
we can write:
\begin{align}
B_{\{2l,n\}}^{fin}\left(p^{2},m_{1}^{2},m_{2}^{2}\right) & =b_{\{2l,n\}}^{fin}\left(1+\epsilon\ln\frac{\mu^{2}}{m_{2}^{2}}\right)+\left(-1\right)^{n}\epsilon\left(d_{1l}I_{1}+d_{2l}I_{2}+d_{3l}I_{3}+d_{3l}I_{1}\ln^{2}\frac{\mu^{2}}{m_{2}^{2}}\right)\nonumber \\
\nonumber \\
\textrm{where}\label{eq:3aa}\\
I_{1} & =\intop_{0}^{1}dx\,x^{n}A^{l}\left(p^{2},m_{1}^{2},m_{2}^{2}\right)\nonumber \\
\nonumber \\
I_{2} & =\intop_{0}^{1}dx\,x^{n}A^{l}\left(p^{2},m_{1}^{2},m_{2}^{2}\right)\ln\frac{m_{2}^{2}}{A\left(p^{2},m_{1}^{2},m_{2}^{2}\right)}\nonumber \\
\nonumber \\
I_{3} & =\intop_{0}^{1}dx\,x^{n}A^{l}\left(p^{2},m_{1}^{2},m_{2}^{2}\right)\ln^{2}\frac{m_{2}^{2}}{A\left(p^{2},m_{1}^{2},m_{2}^{2}\right)}\nonumber \\
\nonumber \\
A\left(p^{2},m_{1}^{2},m_{2}^{2}\right) & =p^{2}x^{2}+m_{1}^{2}+x\left(m_{2}^{2}-m_{1}^{2}-p^{2}\right)-i\varepsilon.\nonumber 
\end{align}
The integrals in Eq.\ref{eq:3aa} can be evaluated analytically, but
that can be done later. The coefficients $d_{il}$ are given in the
Tbl.\ref{tbl-a}. 
\begin{table}
\begin{centering}
\begin{tabular}{|c|c|c|c|}
\hline 
$d_{il}$ & $i=1$ & $i=2$ & $i=3$\tabularnewline
\hline 
\hline 
$l=0$ & $\frac{\pi^{2}}{12}$ & $\frac{1}{2}$ & $0$\tabularnewline
\hline 
$l=1$ & $\frac{12+\pi^{2}}{24}$ & $-\frac{1}{2}$ & $\frac{1}{4}$\tabularnewline
\hline 
$l=2$ & $\frac{21+\pi^{2}}{96}$ & $-\frac{3}{16}$ & $\frac{1}{16}$\tabularnewline
\hline 
$l=3$ & $\frac{85+3\pi^{2}}{1728}$ & $-\frac{11}{288}$ & $\frac{1}{96}$\tabularnewline
\hline 
\end{tabular}
\par\end{centering}
\caption{Coefficients $d_{il}$ used in the representation of the linear in
$\epsilon$ term in Eq.\ref{eq:3aa}.}

\label{tbl-a}
\end{table}
The Eq.\ref{eq:3} is only valid if the Schwartz reflection principle
is applicable and the function $B_{\{2l,n\}}^{fin}\left(z,m_{1}^{2},m_{2}^{2}\right)$
(with $z\in\mathbb{C}$) converges to zero as $1/z^{n\geq2}$ when
$z\rightarrow\infty$. These conditions on Eq.\ref{eq:3} applicability
often require the use of multiple subtractions at a given pole, which
results in replacement of Eq.\ref{eq:3} by the multiply-subtracted
dispersive integral. In our view, the best way to transform Eq.\ref{eq:3}
into the multiply-subtracted dispersive integral is to follow the
same idea as if we would to remove UV-part of Eq.\ref{eq:1} by using
the subtractive scheme at an arbitrary scale $\Lambda$. Of course,
the final result should not depend on any scale, and hence where will
be an additional terms to remove any dependence. To remove the UV-part
of Eq.\ref{eq:1}, we can easily generalize this procedure by using
the following subtractions:
\begin{align}
B_{\{2l,n\}}^{sub}\left(p^{2},m_{1}^{2},m_{2}^{2},\Lambda^{2}\right) & =B_{\{2l,n\}}-\sum_{i=0}^{l}\frac{1}{i!}\left.\left(\frac{\partial^{i}B_{\{2l,n\}}}{\partial\left(p^{2}\right)^{i}}\right)\right|_{p^{2}=\Lambda^{2}}\left(p^{2}-\Lambda^{2}\right)^{i}.\label{eq:4}
\end{align}
Here, $B_{\{2l,n\}}\equiv B_{\{2l,n\}}\left(p^{2},m_{1}^{2},m_{2}^{2}\right)$
and $B_{\{2l,n\}}^{sub}\left(p^{2},m_{1}^{2},m_{2}^{2},\Lambda^{2}\right)$
is multiply-subtracted Eq.\ref{eq:1}. Now, we will subtract and add
the finite part of the second term of Eq.\ref{eq:4} to Eq.\ref{eq:3},
and use the subtracted terms to construct the multiply-subtracted
dispersive integral of Eq.\ref{eq:3}. As a result, we can write the
following:
\begin{align}
 &  & B_{\{2l,n\}}^{fin}\left(p^{2},m_{1}^{2},m_{2}^{2}\right)=\frac{\left(p^{2}-\Lambda^{2}\right)^{l+1}}{\pi}\intop_{\left(m_{1}+m_{2}\right)^{2}}^{\infty}ds\frac{\Im B_{\{2l,n\}}^{fin}\left(s,m_{1}^{2},m_{2}^{2}\right)}{\left(s-p^{2}-i\varepsilon\right)\left(s-\Lambda^{2}-i\varepsilon\right)^{l+1}}\nonumber \\
\label{eq:5}\\
 &  & +\sum_{i=0}^{l}\frac{1}{i!}\left.\left(\frac{\partial^{i}B_{\{2l,n\}}^{fin}\left(p^{2},m_{1}^{2},m_{2}^{2}\right)}{\partial\left(p^{2}\right)^{i}}\right)\right|_{p^{2}=\Lambda^{2}}\left(p^{2}-\Lambda^{2}\right)^{i}.\nonumber 
\end{align}
Eq.\ref{eq:5} has no dependence on the scale $\Lambda$ and its second
term is finite with a polynomial structure in $p^{2}$, which can
be easily evaluated in the second-loop integration. Finally, we can
write dimensionally regularized sub-loop insertion as:
\begin{align}
 & B_{\{2l,n\}}\left(p^{2},m_{1}^{2},m_{2}^{2}\right)=\nonumber \\
\nonumber \\
 & \sum_{i=0}^{l}\Biggl[\biggl(\frac{1}{\epsilon}+\ln\frac{\mu^{2}}{m_{2}^{2}}\biggl)a_{i}^{\{2l,n\}}p^{2i}+\frac{1}{i!}\left.\left(\frac{\partial^{i}B_{\{2l,n\}}^{fin}\left(p^{2},m_{1}^{2},m_{2}^{2}\right)}{\partial\left(p^{2}\right)^{i}}\right)\right|_{p^{2}=\Lambda^{2}}\left(p^{2}-\Lambda^{2}\right)^{i}\Biggl]\label{eq:6}\\
\nonumber \\
 & +\frac{\left(p^{2}-\Lambda^{2}\right)^{l+1}}{\pi}\intop_{\left(m_{1}+m_{2}\right)^{2}}^{\infty}ds\frac{\Im B_{\{2l,n\}}^{fin}\left(s,m_{1}^{2},m_{2}^{2}\right)}{\left(s-p^{2}-i\varepsilon\right)\left(s-\Lambda^{2}-i\varepsilon\right)^{l+1}}.\nonumber 
\end{align}
The first term of Eq.\ref{eq:6} will contribute to the numerator
algebra and the second term will add an additional propagator $\frac{\left(p^{2}-\Lambda^{2}\right)^{l+1}}{s-p^{2}-i\varepsilon}$
to the second-loop integral. 

In the case of the triangle insertion, the three-point PV functions
which can be written in the form of the derivatives of the two-point
functions. To begin with, the scalar three-point function function
is given by:
\begin{align}
 & C_{0}\equiv C_{0}\left(p_{1}^{2},p_{2}^{2},\left(p_{1}+p_{2}\right)^{2},m_{1}^{2},m_{2}^{2},m_{3}^{2}\right)=\nonumber \\
\label{eq:7}\\
 & \frac{\mu^{4-D}}{i\pi^{D/2}}\int d^{D}q\frac{1}{\left[q^{2}-m_{1}^{2}\right]\left[\left(q+p_{1}\right)^{2}-m_{2}^{2}\right]\left[\left(q+p_{1}+p_{2}\right)^{2}-m_{3}^{2}\right]}.\nonumber 
\end{align}
With Feynman's trick, we can join the first two propagators in Eq.\ref{eq:7},
and after shifting momentum $q=\tau-p_{1}-p_{2}$, we can write:
\begin{align}
 & C_{0}=\frac{\mu^{4-D}}{i\pi^{D/2}}\intop_{0}^{1}dx\int d^{D}\tau\frac{1}{\left[\left(\tau-\left(p_{1}\bar{x}+p_{2}\right)\right)^{2}-m_{12}^{2}\right]^{2}\left[\tau^{2}-m_{3}^{2}\right]}\nonumber \\
\label{eq:8}\\
 & m_{12}^{2}=m_{1}^{2}\bar{x}+m_{2}^{2}x-p_{1}^{2}x\bar{x}.\nonumber 
\end{align}
Here, $\bar{x}=1-x$, and momentum $p_{1}$ does not enter the second
loop integral and is treated as a combination of the external momenta
of the two-loop graph. Term $\left(\left(\tau-\left(p_{1}\bar{x}+p_{2}\right)\right)^{2}-m_{12}^{2}\right)^{-2}$
can be replaced after shifting mass $m_{12}^{2}$ by a small parameter
$\phi$: 
\begin{align}
\frac{1}{\left(\left(\tau-\left(p_{1}\bar{x}+p_{2}\right)\right)^{2}-m_{12}^{2}\right)^{2}} & =\underset{\phi\rightarrow0}{\lim}\frac{\partial}{\partial\phi}\left[\frac{1}{\left(\tau-\left(p_{1}\bar{x}+p_{2}\right)\right)^{2}-\left(m_{12}^{2}+\phi\right)}\right].\label{eq:9}
\end{align}
As a result, Eq.\ref{eq:8} can be represented in the form of

\begin{align}
 & C_{0}=\frac{\mu^{4-D}}{i\pi^{D/2}}\lim_{\phi\rightarrow0}\frac{\partial}{\partial\phi}\intop_{0}^{1}dx\int d^{D}\tau\frac{1}{\left[\left(\tau-\left(p_{1}\bar{x}+p_{2}\right)\right)^{2}-\left(m_{12}^{2}+\phi\right)\right]\left[\tau^{2}-m_{3}^{2}\right]}=\nonumber \\
\label{eq:10}\\
 & \lim_{\phi\rightarrow0}\frac{\partial}{\partial\phi}\intop_{0}^{1}dx\;B_{0}\left(\left(p_{1}\bar{x}+p_{2}\right)^{2},m_{3}^{2},m_{12}^{2}+\phi\right).\nonumber 
\end{align}
Since $C_{0}$ function is UV finite, its dispersive representation
will be given by a singly subtracted integral:
\begin{align}
C_{0} & =\lim_{\phi\rightarrow0}\frac{\partial}{\partial\phi}\intop_{0}^{1}dx\;\Biggl[\ln\frac{m_{3}^{2}}{m_{12}^{2}+\phi}+B_{0}^{fin}\left(\Lambda^{2},m_{3}^{2},m_{12}^{2}+\phi\right)\nonumber \\
\label{eq:11}\\
 & +\frac{\left(\left(p_{1}\bar{x}+p_{2}\right)^{2}-\Lambda^{2}\right)}{\pi}\intop_{\left(m_{3}+\sqrt{m_{12}^{2}+\phi}\right)^{2}}^{\infty}ds\frac{\Im B_{0}^{fin}\left(s,m_{3}^{2},m_{12}^{2}+\phi\right)}{\left(s-\left(p_{1}\bar{x}+p_{2}\right)^{2}-i\varepsilon\right)\left(s-\Lambda^{2}-i\varepsilon\right)}\Biggr].\nonumber 
\end{align}
In this representation of $C_{0}$ function, we have momentum $p_{2}$
as a combination of the second-loop and external momenta. When taking
a derivative with respect to the mass shift parameter $\phi$, we
use transformation $\ln\frac{\mu^{2}}{m_{12}^{2}+\phi}\rightarrow\ln\frac{\mu^{2}}{m_{3}^{2}}+\ln\frac{m_{3}^{2}}{m_{12}^{2}+\phi}$
in order to remove $\mu$-scale dependence from the Feynman integral.
The finite part of the $B_{0}$ function has a rather simple analytical
structure: 
\begin{align}
B_{0}^{fin}\left(p^{2},m_{1}^{2},m_{2}^{2}\right) & =2+\frac{\kappa^{1/2}\left(p^{2},m_{1}^{2},m_{2}^{2}\right)}{p^{2}}\ln\left(\frac{\kappa^{1/2}\left(p^{2},m_{1}^{2},m_{2}^{2}\right)+m_{1}^{2}+m_{2}^{2}-p^{2}}{2m_{1}m_{2}}\right)\nonumber \\
\label{eq:12}\\
 & -\frac{\left(m_{1}^{2}-m_{2}^{2}+p^{2}\right)}{2p^{2}}\ln\left(\frac{m_{1}^{2}}{m_{2}^{2}}\right).\nonumber 
\end{align}
Here, $\kappa\left(p^{2},m_{1}^{2},m_{2}^{2}\right)$ is a Källen
function, $\kappa\left(a,b,c\right)=a^{2}+b^{2}+c^{2}-2\left(ab+bc+ac\right)$.
In the case of the higher rank three-point tensor coefficient functions,
we can represent them through a combinations of $B_{\{2l,n\}}$ functions
following the prescription of \citep{AA1}:
\begin{align}
 & C_{\underset{2l}{\underbrace{0...0}}\,\underset{n}{\underbrace{1...1}}\,\underset{m}{\underbrace{2...2}}}\equiv C_{\{2l,n,m\}}=\lim_{\phi\rightarrow0}\frac{\partial}{\partial\phi}\intop_{0}^{1}dx\,x^{n}\sum_{i=0}^{m}b_{i}^{\{m\}}B_{\{2l,i+n\}}.\label{eq:13}
\end{align}
Here, $B_{\{2l,i+n\}}\equiv B_{\{2l,i+n\}}\left(\left(p_{1}\bar{x}+p_{2}\right)^{2},m_{3}^{2},m_{12}^{2}+\phi\right)$,
and the UV-divergent three-point functions have $l\geqslant1$. Coefficients
$b_{i}^{\{m\}}$ are given in the Tbl.\ref{tbl:2}. 
\begin{table}
\begin{centering}
\begin{tabular}{|c|c|c|c|c|c|}
\hline 
$b_{i}^{\{m\}}$ & $i=0$ & $i=1$ & $i=2$ & $i=3$ & $i=4$\tabularnewline
\hline 
\hline 
$m=0$ & $1$ & $\ldots$ & $\ldots$ & $\ldots$ & $\ldots$\tabularnewline
\hline 
$m=1$ & $-1$ & $-1$ & $\ldots$ & $\ldots$ & $\ldots$\tabularnewline
\hline 
$m=2$ & $1$ & $2$ & $1$ & $\ldots$ & $\ldots$\tabularnewline
\hline 
$m=3$ & $-1$ & $-3$ & $-3$ & $-1$ & $\ldots$\tabularnewline
\hline 
$m=4$ & $1$ & $4$ & $6$ & $4$ & $1$\tabularnewline
\hline 
\end{tabular}
\par\end{centering}
\caption{Expansion coefficients $b_{i}^{\{m\}}$for many-points Passarino-Veltman
functions.}

\label{tbl:2}
\end{table}
Using Eq.\ref{eq:6} in Eq.\ref{eq:13}, we can write the generalized
three-point function dispersively with dimensionally regularized UV-divergence:
\begin{align}
 & C_{\{2l,n,m\}}=\lim_{\phi\rightarrow0}\frac{\partial}{\partial\phi}\intop_{0}^{1}dx\,x^{n}\sum_{i=0}^{m}b_{i}^{\{m\}}\Biggl(\sum_{j=0}^{l}\Biggl[\biggl(\frac{1}{\epsilon}+\ln\frac{\mu^{2}}{m_{12}^{2}+\phi}\biggl)a_{j}^{\{2l,i+n\}}p_{12x}^{2j}\nonumber \\
\nonumber \\
 & +\frac{1}{j!}\left.\biggl(\frac{\partial^{j}B_{\{2l,i+n\}}^{fin}\left(p^{2},m_{3}^{2},m_{12}^{2}+\phi\right)}{\partial\left(p^{2}\right)^{j}}\biggl)\right|_{p^{2}=\Lambda^{2}}\left(p_{12x}^{2}-\Lambda^{2}\right)^{j}\Biggr]\nonumber \\
\label{eq:13a}\\
 & +\frac{\left(p_{12x}^{2}-\Lambda^{2}\right)^{l+1}}{\pi}\intop_{\left(m_{3}+\sqrt{m_{12}^{2}+\phi}\right)^{2}}^{\infty}ds\frac{\Im B_{\{2l,i+n\}}^{fin}\left(s,m_{3}^{2},m_{12}^{2}+\phi\right)}{\left(s-p_{12x}^{2}-i\varepsilon\right)\left(s-\Lambda^{2}-i\varepsilon\right)^{l+1}}\Biggr),\nonumber 
\end{align}
with $p_{12x}$ is defined as $p_{12x}=p_{1}\bar{x}+p_{2}$. As an
example, let's consider expression for $C_{001}$ where UV-divergent
pole is extracted explicitly:
\begin{align}
 & C_{001}=-\frac{1}{12}\left(\frac{1}{\epsilon}+\ln\frac{\mu^{2}}{m_{3}^{2}}\right)\nonumber \\
\nonumber \\
 & +\lim_{\phi\rightarrow0}\frac{\partial}{\partial\phi}\intop_{0}^{1}dx\,x\Biggl(\frac{1}{12}\left(\frac{1}{2}p_{12x}^{2}-m_{3}^{2}-2\left(m_{12}^{2}+\phi\right)\right)\ln\frac{m_{3}^{2}}{m_{12}^{2}+\phi}\nonumber \\
\label{eq:13b}\\
 & +B_{001}^{fin}\left(\Lambda^{2},m_{3}^{2},m_{12}^{2}+\phi\right)+\left.\biggl(\frac{\partial B_{001}^{fin}\left(p^{2},m_{3}^{2},m_{12}^{2}+\phi\right)}{\partial p^{2}}\biggl)\right|_{p^{2}=\Lambda^{2}}\left(p_{12x}^{2}-\Lambda^{2}\right)\nonumber \\
\nonumber \\
 & +\frac{\left(p_{12x}^{2}-\Lambda^{2}\right)^{2}}{\pi}\intop_{\left(m_{3}+\sqrt{m_{12}^{2}+\phi}\right)^{2}}^{\infty}ds\frac{\Im B_{001}^{fin}\left(s,m_{3}^{2},m_{12}^{2}+\phi\right)}{\left(s-p_{12x}^{2}-i\varepsilon\right)\left(s-\Lambda^{2}-i\varepsilon\right)^{2}}\Biggr).\nonumber 
\end{align}
 To derive expressions for the four-point PV functions in the two-point
function basis, we can use the ideas outlined in Eqns.\ref{eq:7}-\ref{eq:10}:
\begin{align}
 & D_{\underset{2l}{\underbrace{0...0}}\,\underset{n}{\underbrace{1...1}}\,\underset{k}{\underbrace{2...2}}\,\underset{m}{\underbrace{3...3}}}\equiv D_{\{2l,n,k,m\}}=\lim_{\phi\rightarrow0}\frac{\partial^{2}}{\partial\phi^{2}}\intop_{0}^{1}dx\,x^{n}\intop_{0}^{1-x}dy\,y^{k}\sum_{i=0}^{m}b_{i}^{\{m\}}B_{\{2l,i+n+k\}},\label{eq:14}
\end{align}
where $D_{\{2l,n,k,m\}}\equiv D_{\{2l,n,k,m\}}\left(p_{1}^{2},p_{2}^{2},p_{3}^{2},p_{4}^{2},\left(p_{1}+p_{2}\right)^{2},\left(p_{2}+p_{3}\right)^{2},m_{1}^{2},m_{2}^{2},m_{3}^{2},m_{4}^{2}\right)$
and $B_{\{2l,i+n+k\}}\equiv B_{\{2l,i+n+k\}}\left[\left(p_{1}\left(\bar{x}-y\right)+p_{2}\bar{y}+p_{3}\right)^{2},m_{4}^{2},m_{123}^{2}+\phi\right]$
with $m_{123}^{2}=m{}_{1}^{2}\left(\bar{x}-y\right)+m_{2}^{2}x+m_{3}^{2}y-p_{1}^{2}x\bar{x}-p_{12}^{2}y\bar{y}+2xy\left(p_{1}p_{12}\right)$
and $p_{12}=p_{1}+p_{2}$. As a result, the dispersive generalization
can be written as:
\begin{align}
 & D_{\{2l,n,k,m\}}=\lim_{\phi\rightarrow0}\frac{\partial^{2}}{\partial\phi^{2}}\intop_{0}^{1}dx\,x^{n}\intop_{0}^{1-x}dy\,y^{k}\sum_{i=0}^{m}b_{i}^{\{m\}}\Biggl(\sum_{j=0}^{l}\Biggl[\biggl(\frac{1}{\epsilon}+\ln\frac{\mu^{2}}{m_{123}^{2}+\phi}\biggl)a_{j}^{\{2l,i+n+k\}}p_{123xy}^{2j}\nonumber \\
\nonumber \\
 & +\frac{1}{j!}\left.\biggl(\frac{\partial^{j}B_{\{2l,i+n+k\}}^{fin}\left(p^{2},m_{4}^{2},m_{123}^{2}+\phi\right)}{\partial\left(p^{2}\right)^{j}}\biggl)\right|_{p^{2}=\Lambda^{2}}\left(p_{123xy}^{2}-\Lambda^{2}\right)^{j}\Biggr]\label{eq:14a}\\
\nonumber \\
 & +\frac{\left(p_{123xy}^{2}-\Lambda^{2}\right)^{l+1}}{\pi}\intop_{\left(m_{4}+\sqrt{m_{123}^{2}+\phi}\right)^{2}}^{\infty}ds\frac{\Im B_{\{2l,i+n+k\}}^{fin}\left(s,m_{4}^{2},m_{123}^{2}+\phi\right)}{\left(s-p_{123xy}^{2}-i\varepsilon\right)\left(s-\Lambda^{2}-i\varepsilon\right)^{l+1}}\Biggr).\nonumber 
\end{align}
Here, we have $p_{123xy}=p_{1}\left(\bar{x}-y\right)+p_{2}\bar{y}+p_{3}$.
Eq.\ref{eq:14a} shows that the UV-divergent four-point functions
show up at $l\geqslant2$. The five-point function also can be easily
expressed in two-point function basis:
\begin{align}
 & E_{\underset{2l}{\underbrace{0...0}}\,\underset{n}{\underbrace{1...1}}\,\underset{k}{\underbrace{2...2}}\,\underset{r}{\underbrace{3...3}}\,\underset{m}{\underbrace{4...4}}}\equiv E_{\{2l,n,k,r,m\}}=\nonumber \\
\label{eq:15}\\
 & \lim_{\phi\rightarrow0}\frac{\partial^{3}}{\partial\phi^{3}}\intop_{0}^{1}dx\,x^{n}\intop_{0}^{1-x}dy\,y^{k}\intop_{0}^{1-x-y}dz\,z^{r}\sum_{i=0}^{m}b_{i}^{\{m\}}B_{\{2l,i+n+k+r\}}.\nonumber 
\end{align}
Here, $E_{\{2l,n,k,r,m\}}\equiv E_{\{2l,n,k,r,m\}}\left(p_{1}^{2},p_{2}^{2},p_{3}^{2},p_{4}^{2},p_{5}^{2},p_{12}^{2},p_{23}^{2},p_{34}^{2},p_{45}^{2},p_{51}^{2},m_{1}^{2},m_{2}^{2},m_{3}^{2},m_{4}^{2},m_{5}^{2}\right)$
with $p_{ij}=\left(p_{i}+p_{j}\right)^{2}$, $p_{ijk}=\left(p_{i}+p_{j}+p_{k}\right)^{2}$,
and $B_{\{2l,i+n+k+r\}}\equiv B_{\{2l,i+n+k+r\}}\left(\left(p_{1}\left(\bar{x}-y-z\right)+p_{2}\left(\bar{y}-z\right)+p_{3}\bar{z}+p_{4}\right)^{2},m_{5}^{2},m_{1234}^{2}+\phi\right)$
with $m_{1234}^{2}=m_{1}^{2}\left(\bar{x}-y-z\right)+m_{2}^{2}x+m_{3}^{2}y+m_{4}^{2}z-p_{1}^{2}\bar{x}x-p_{12}^{2}\bar{y}y-p_{123}^{2}\bar{z}z+2xy\left(p_{1}p_{12}\right)+2xz\left(p_{1}p_{123}\right)+2yz\left(p_{12}p_{123}\right)$.
The dispersive generalization of the five-point function is given
in a similar way:
\begin{align}
 & E_{\{2l,n,k,r,m\}}=\lim_{\phi\rightarrow0}\frac{\partial^{3}}{\partial\phi^{3}}\intop_{0}^{1}dx\,x^{n}\intop_{0}^{1-x}dy\,y^{k}\intop_{0}^{1-x-y}dz\,z^{r}\nonumber \\
\nonumber \\
 & \times\sum_{i=0}^{m}b_{i}^{\{m\}}\Biggl(\sum_{j=0}^{l}\Biggl[\biggl(\frac{1}{\epsilon}+\ln\frac{\mu^{2}}{m_{1234}^{2}+\phi}\biggl)a_{j}^{\{2l,i+n+k+r\}}p_{1234xyz}^{2j}\nonumber \\
\label{eq:15a}\\
 & +\frac{1}{j!}\left.\biggl(\frac{\partial^{j}B_{\{2l,i+n+k+r\}}^{fin}\left(p^{2},m_{5}^{2},m_{1234}^{2}+\phi\right)}{\partial\left(p^{2}\right)^{j}}\biggl)\right|_{p^{2}=\Lambda^{2}}\left(p_{1234xyz}^{2}-\Lambda^{2}\right)^{j}\Biggr]\nonumber \\
\nonumber \\
 & +\frac{\left(p_{1234xyz}^{2}-\Lambda^{2}\right)^{l+1}}{\pi}\intop_{\left(m_{5}+\sqrt{m_{1234}^{2}+\phi}\right)^{2}}^{\infty}ds\frac{\Im B_{\{2l,i+n+k+r\}}^{fin}\left(s,m_{5}^{2},m_{1234}^{2}+\phi\right)}{\left(s-p_{1234xyz}^{2}-i\varepsilon\right)\left(s-\Lambda^{2}-i\varepsilon\right)^{l+1}}\Biggr),\nonumber 
\end{align}
where momentum $p_{1234xyz}$ is defined as $p_{1234xyz}=p_{1}\left(\bar{x}-y-z\right)+p_{2}\left(\bar{y}-z\right)+p_{3}\bar{z}+p_{4}$.

\section{Conclusion}

In this work, we have extracted the UV-divergent poles of the Passarino-Veltman
functions analytically and presented them as the dimensionally-regularized
and multiply-subtracted dispersive sub-loop insertions. We have also
retained the terms linear in $\epsilon$, which are required to produce
local terms for the second-loop integration. Finally, all sub-loop
insertions are conveniently expressed in the two-point function basis,
which allows to carry out the calculations analytically, with numerical
integration done only over the Feynman and dispersion parameters.
As a result, this approach will allow to speed up calculations for
the two-loop radiative corrections and to better account for the experiment-specific
kinematics.
\begin{acknowledgments}
The authors are grateful to A. Davydychev, H. Spiesberger and M. Vanderhaeghen
for the fruitful and exciting discussions. We would also like to express
special thanks to the Institut für Kernphysik of Johannes Gutenberg-Universität
Mainz for hospitality and support. This work was funded by the Natural
Sciences and Engineering Research Council (NSERC) of Canada. 
\end{acknowledgments}

\end{document}